\documentstyle[12pt]{article}
\begin{document}

{\footnotesize {hep-th/9608061 {\hfill } USC-96/HEP-B5}} \bigskip

\begin{center}
{\normalsize {\bf ALGEBRAIC\ STRUCTURE\ OF\ S-THEORY}}\footnote{%
Based on lectures at the Second International Sakharov Conference on
Physics, Moscow, May 1996 and at Strings '96, Santa Barbara, CA, July 1996.}

\baselineskip=22pt
\end{center}

%\vfill
%\vspace*{0.6cm}
\centerline{\footnotesize ITZHAK BARS} \baselineskip=13pt

\centerline{\footnotesize {\it Department of Physics and Astronomy,
University of Southern California}} \baselineskip=12pt \centerline{%
{\footnotesize {\it Los Angeles, CA 90089-0484, USA}}}

\centerline{\footnotesize E-mail: bars@physics.usc.edu}

%\vfill
\vspace*{0.9cm} {\ \centering{%
\begin{minipage}{12.2truecm}\footnotesize\baselineskip=12pt\noindent
\centerline{\footnotesize ABSTRACT}\vspace*{0.3cm}
\parindent=0pt \
The algebraic structure of S-Theory and its representations are described. 
This structure includes up to 13 hidden dimensions. It implies the existence 
of an SO(10,2) covariant supergravity theory
as a limit of the secret theory behind string theory.
The black hole entropy is invariant under transformations including 
the 11th and 12th dimensions. The discussion includes generalization into 
curved spacetime via various contractions of the superalgebras OSp(1/32), OSp(1/64) etc.
\end{minipage}}}

%\vspace*{0.6cm}

\baselineskip=15pt \setcounter{footnote}{0} \renewcommand{\thefootnote}{%
\alph{footnote}}

%%%%%%%%%%%%%%%%%%%

\section{Introduction}

Ever since Witten's paper in Strings'95 \cite{witten} there has been a lot
of discussion on dualities \cite{ht} and hidden dimensions \cite{witten}-%
\cite{martinec} as properties of an underlying secret theory that includes
superstring theory and its super p-brane { and D-brane \cite{pol}
generalizations. The secret theory has been called M-theory \cite{jhs},
F-theory \cite{vafaf}, S-theory \cite{ibstheory}, Y-theory \cite{hullyth}
... where each version emphasizes some aspects of the underlying theory. In
this talk I will describe the higher algebraic structures that define
S-theory.

S-theory \cite{ibstheory}\cite{ibbeyond} concentrates on certain global
aspects of the secret theory based on a superalgebra that is common to all
the incarnations of the secret theory. It provides an algebraic approach to
study the secret theory through representation theory of the superalgebra.
Dualities, hidden dimensions p-branes etc. are some of the properties that
are manifest in S-theory.

The superalgebra has 32 fermionic and 528 bosonic generators. The maximum
number of supercharges in a physical theory is 32. This constraint comes
from 4-dimensions, which admits at the most 8 supercharges, since there can
be no supermultiplet of interacting massless particles with spins higher
than 2. The collection of all bosonic operators form a 32$\times 32$
symmetric matrix $S$ (528 hermitian generators) given by the anticommutator
of the supercharges $\left\{ Q,Q\right\} \sim S$. The representations of the
superalgebra are intimately connected to the properties of $S.$ The
structure and symmetries of $S$ are related to p-branes, dualities and
hidden dimensions. Global properties, certain states and certain
non-perturbative properties of the underlying theory may be studied by
analyzing the representations of this superalgebra. This line of
investigation is called S-theory.

The 528 bosonic generators include momenta $P_\mu $ and central extensions.
The Lorentz scalar central extensions correspond to charges of zero-branes, i.e.
particles. The Lorentz non-scalar central extensions $%
Z_{\mu _1\cdots \mu _p}^{ab}$ are related to boundaries of p-branes (for
example for an open string, p=1, $Z_{\mu _1}=X_{\mu _1}(0)-X_{\mu _1}(\pi )$
is the difference of end point positions \cite{ibbeyond}). In the simplest
form of S-theory p-branes propagate in flat space. Then all 528 generators
commute with each other and they can be simultaneously diagonalized. Central
extensions are then on an equal footing with momenta, and therefore
``momentum space'' is enlarged in the secret theory. One may think of this
as a generalization of the concept of flat spacetime. The representations of
the superalgebra are then easy to construct. The space of states include
long supermultiplets as well as shorter multiplets of various kinds,
corresponding to BPS-like states \cite{ibstheory}. Algebraically, the flat
version of S-theory corresponds to a contraction of the superalgebra
OSp(1/32) such that all bosonic generators commute. This is achieved by 
rescaling the OSp(1/32) bosonic generators
and then taking the contraction limit.

In this talk I will concentrate mainly on S-theory in flat spacetime. At the
end, I will indicate possible directions that may be pursued to consider
S-theory in curved compactified spaces or curved spacetime, and the
relevance of other contractions of OSp(1/32), OSp(1/64), etc...

\section{The extended superalgebra}

In an arbitrary number of dimensions, we label the 32 supercharges as 
\begin{equation}
Q_\alpha ^a=\left\{ 
\begin{array}{l}
a=1,2,\cdots N;\quad \rm{spinor\ in\ }$c+2$\rm{\ dims.} \\ 
\alpha \,\,\rm{=spinor\ in\ }$d$\rm{\ dims}.
\end{array}
\right.
\end{equation}
Here $N=1$, in $d$=11; $N=2,$ in $d$=10;$\cdots\ ,\ N=8$, in $d$=4, and 
$c$=number of compactified string dimensions, such that $d+c=10.$ By
interpreting $a$ as the label of a spinor in $c+2$ dimensions we begin to
see that there are at least 2 hidden dimensions. $a$ is also a label for an
irreducible representation of the group $K,$ where $K$ is the maximal
compact subgroup of $U$-duality. For example for $d=4,\,\,c=6,$ $\,N=8$, the
index $a$ is interpreted either as the complex {\bf 8 (}or {\bf 8*) }of $%
K=SU(8)\subset E_{7,7}$ or as the spinors {\bf 8}$_{\pm }$ of $%
SO(c+1,1)=SO(7,1).$ The $K$ classification gives information about dualities
while the $SO(c+1,1)$ classification gives information about the hidden
dimensions \cite{ibbeyond} \cite{ibstheory}. 
The 528 generators as well as all the states of the theory can
be classified in either way. The two classsifications are not contained in
each other. If one classifies according to duality one looses track of the
hidden dimensions and vice versa. Completeness implies that the collection
of representations in one classification can be expanded into a collection
of representations in the other classification. This observation can be used
as a tool for finding some of the states of the secret theory by starting
from the well known string states \cite{ib11}\cite{ibbeyond}.

The extended superalgebra in d-dimensions has the form 
\begin{equation}
\begin{array}{c}
\left\{ Q_\alpha ^a,Q_\beta ^b\right\} =\left( S\right) _{\alpha \beta }^{ab}
\\ 
\left( S\right) _{\alpha \beta }^{ab}=\delta ^{ab}\gamma _{\alpha \beta
}^\mu \,\,P_\mu +\sum_{p=0,1,,2,\cdots }\gamma _{\alpha \beta }^{\mu
_1\cdots \mu _p}\,\,\,Z_{\mu _1\cdots \mu _p}^{ab}.
\end{array}
\end{equation}
where $P_\mu $ is the momentum operator and $\,Z_{\mu _1\cdots \mu _p}^{ab}$
are the central extensions. It can be shown that (for $d\leq $9) this can be
derived by compactifying either a type-A superalgebra in 12D or a type-B
superalgebra in 10D$\otimes $3D, both embedded in the spinor space of 13D 
\cite{ibstheory}. Hence S-theory is related to both M-theory (A-type,11D)
and F-theory (B-type, 12D$\rightarrow $10D). The spinor space of 13D with
(11,2) signature has 64 spinors. There are two distinct projections that
have 32 real spinors. The type-A projection distinguishes the 13th
dimension, leaving an SO(10,2) covariance. The type-B projection
distinguishes 10D from 3D, leaving an SO(9,1)$\otimes $SO(2,1)  
covariance\footnote{
One may be tempted to distinguish the 0'-direction instead of the 13th
and get a projection with (11,1) signature, but this projection gives
a complex 32-component spinor, which is really equivalent to 64 supercharges.
This would contradict the basic theorem in 4D that prevents more than
32 supercharges.}.
Under compactification, the type-A(B) reduces to the type-A(B) superalgebra
of type-IIA(B) string. The geometrical origin of the SL(2,R) duality
symmetry of type-IIB string is the extra 3D. The T-duality symmetry between
types A,B involves the mixing of the 13th dimension with the others. So
there are traces of up to 13 hidden dimensions. Note that it is the spinor
space of 13D that is relevant so far. As emphasized in the introduction, the
meaning of ``dimensions'' and of ``spacetime'' get generalized in unusual
ways (not just 13D, but 528 bosonic generators), and as noted below, there
is only one time coordinate corresponding to the trace of S \cite{ibstheory}.

In 12D the extended superalgebra takes the form \cite{ibbeyond}

\begin{equation}
\left\{ Q_\alpha ,Q_\beta \right\} =\gamma _{\alpha \beta
}^{M_1M_2}\,\,\,Z_{M_1M_2}+\gamma _{\alpha \beta }^{M_1\cdots
M_6}\,\,\,Z_{M_1\cdots M_6}^{+}  \label{type2}
\end{equation}
where $M=0^{\prime },\mu ,$ and $\mu =0,1,2,\cdots 10,$ with two time-like
dimensions denoted by $M=0^{\prime },$ $0.$ The six-index tensor is self
dual in 12D. By compactifying the $0^{\prime }$ coordinate the 11D extended
superalgebra \cite{townsend} emerges 
\begin{equation}
\begin{array}{l}
Z_{M_1M_2}\rightarrow P_\mu \oplus Z_{\mu _1\mu _2}\quad 66=11+55 \\ 
Z_{M_1\cdots M_6}^{+}\rightarrow X_{\mu _1\cdots \mu _5}\quad \quad \quad
462=462
\end{array}
\end{equation}

Note that there is no translation operator $P_M$ in the (10,2) version.
Therefore the extension of the theory from (10,1) to (10,2) is not the naive
extension that would have implied two time coordinates, since the
corresponding canonical conjugate momenta are not present. There is only one
time translation operator, hence there is only one time coordinate that can
be recognized only in the 11D (or lower D) notation.

Another remark is that the 12D Lorentz generator $L_{MN}$ does not appear on
the right hand side; $Z_{M_1M_2}$ or $Z_{\mu _1\mu _2}^{ab}$ are not related
to $L_{MN}.$ The fermionic or bosonic generators of the superalgebra do not
commute with $L_{MN}$ since they are classified as spinors or tensors under
Lorentz transformations. Hence, including $L_{MN}$ would enlarge the algebra
further in a trivial form.

The type-B superalgebra in 10D$\otimes $3D can be written as 
\begin{equation}
\begin{array}{c}
\left\{ Q_{\bar{\alpha}}^{\bar{a}},Q_{\bar{\beta}}^{\bar{b}}\right\} =\left(
S_B\right) _{\bar{\alpha}\bar{\beta}}^{\bar{a}\bar{b}} \\ 
\left( S_B\right) _{\bar{\alpha}\bar{\beta}}^{\bar{a}\bar{b}}=\bar{\gamma}_{%
\bar{\alpha}\bar{\beta}}^{\bar{\mu}}\,\,\left( c\bar{\tau}_i\right) ^{\bar{a}%
\bar{b}}P_{\bar{\mu}}^i+\bar{\gamma}_{\bar{\alpha}\bar{\beta}}^{\bar{\mu}_1%
\bar{\mu}_2\bar{\mu}_3}\,\,c^{\bar{a}\bar{b}}\,Y_{\bar{\mu}_1\bar{\mu}_2\bar{%
\mu}_3}+\bar{\gamma}_{\bar{\alpha}\bar{\beta}}^{\bar{\mu}_1\cdots \bar{\mu}%
_5}\,\,\,\left( c\bar{\tau}_i\right) ^{\bar{a}\bar{b}}X_{\bar{\mu}_1\cdots 
\bar{\mu}_5}^i.
\end{array}
\label{type2b}
\end{equation}
where $\bar{\alpha},\bar{\beta}=1,2,\cdots ,16$ and $\bar{a},\bar{b}=1,2$
while $\bar{\mu}=0,1,\cdots ,9$ and $i=0^{\prime },1^{\prime },2^{\prime }.$
The $X_{\mu _1\cdots \mu _5}^i$ are self dual in 10D. Here $\bar{\gamma}_{%
\bar{\alpha}\bar{\beta}}^{\bar{\mu}}\,$are 16$\times 16$ 10D gamma matrices
and the $\bar{\tau}_i^{ab}$ are 2$\times 2$ gamma matrices in some hidden 3D
Minkowski space, with $\,c^{ab}=i\sigma _2^{ab}=\varepsilon ^{ab}.$ This
algebra is covariant under SO(1,9)$\otimes SO(1,2).$ Using the SO(11,2)
gamma matrices in \cite{ibstheory} one can see that it can be embedded in 13
dimensions.

All bosonic operators commute with each other. In this sense they all behave
like the momentum operators, and hence can be simultaneously diagonalized.
The type A,B superalgebras are two versions of S-theory that are embedded in
contractions of OSp(1/32) with different choices of bases that have
different isometries.

\section{Representations, BPS states, predictions}

A basic hypothesis is that the superalgebra that defines S-theory is valid
as a dynamical stucture of the secret theory (including broken symmetries).
Then certain global properties of the secret theory can be described
non-perturbatively as properties of the superalgebra itself. For example the
supermultiplet structure of the spectrum of the theory, relations among
correlation functions or scattering amplitudes, coupling constants, etc. can
be computed directly as the properties of the representations of the
superalgebra. In spirit there is a similarity to the theory of current
algebras used in the 60's to explore the properties of strong and weak
interactions, without knowing the details of the underlying theory now known
as the Standard Model.

The physical states of the secret theory (i.e. after gauge freedom and
auxiliary fields have been eliminated) must form multiplets of the
superalgebra consistent with its isometries in various dimensions. For the
flat version of S-theory one can work in a basis in which all 528 bosonic
generators are diagonal (in the more familiar states, such as string states,
most of these eigenvalues vanish, leaving only spacetime momentum,
Kaluza-Klein compactified momentum and winding numbers, as the non-zero
eigenvalues that specify the ``base''). Simultaneously, one can specify a
representation of the isometries which may be reducible (in string theory
this arises by applying oscillators on the base; then at a fixed level there
is a collection of representations of the rotation group and/or other
compactification isometries). Then one can start with a reference state of
the form 
\begin{equation}
|528\rm{\ eigenvalues;\ representation\ of\ isometries\ at\ some\ level\ }l\,\,%
\rm{{>}}  \label{refstate}
\end{equation}
and apply all possible polynomials of the fermionic generators in order to
obtain a supermultiplet. This construction is analogous to the construction of representations
of standard supersymmetry, and hence the familiar supersymmetry multiplets
get generalized by replacing the momentum by the 528 bosonic generators. 
In the generic case one obtains the long
supermultiplet of dimension $2^{32/2}=2_{bosons}^{15}+2_{fermions}^{15}.$
However, under special conditions on the eigenvalues there are shorter
supermultiplets. The condition for shorter multiplets is $detS=0.$ If the
multiplicity of the zero eigenvalue of $S$ is $2n,$ then $2n$ supercharges
vanish on the reference state, and the remaining supercharges act
non-trivially, producing a supermultiplet of dimension $2^{(32-2n)/2}.$ These
are the BPS-type states. In the context of S-theory there is a much richer
spectrum of BPS-type states than the ones discussed previously
in M- or F-theories because of
the inclusion of Lorentz non-singlet central extensions with non-zero
eigenvalues \cite{ibstheory}.

The supermultiplet should include string states of a given level for
consistency with string theory, but in addition, the collection of
representations should be consistent with being expressible as a set of
representations of $K$ (duality) or as a set of representations of $%
SO(c+1,1) $ (hidden dimensions). This tripple constraint is extremely strong
and it typically demands the inclusion of additional states, beyond the
string states of a given level, in order to have a complete supermultiplet.
The additional states are predictions about the secret theory. In some
examples \cite{ib11} they have been interpreted as D-branes \cite{sen-dbr},
but it is not clear that the full interpretation has been given so far.

As an outcome of representation theory one can give some new results that
have not been obtained in other approaches.

\begin{itemize}
\item  It is well known that by keeping only the Lorentz singlet charges one
describes black holes as BPS states. The degeneracy of these states
coincides with the entropy of the black holes. This entropy is is a function
of only the Lorentz singlet central extensions \cite{ferrara} and can be
expressed in terms of natural invariants of U-duality, such as \cite{kallosh}
$E_{7,7}\,(E_{6,6})$ in $d=4$ (5). What was not noticed before is that the
entropy is also invariant under transformations involving hidden dimensions.
This is seen by reclassifying \cite{ibbeyond}
the central extensions as representations of 
$SO(c+1,1)$ that includes two hidden dimensions. Then the blackhole entropy
can be rewritten explicitly as an invariant of $SO(7,1)$ in $d=4$ and $%
SO(6,1)$ in $d=5$, etc.. and therefore provides evidence for 2 hidden
dimensions, one spacelike and the other timelike \cite{ibentropy}. The $%
SO(c+1)$ subgroup is also a subgroup of $K\subset U$, so that it is already
included in U-duality. This part which indicates the presence of the 11th
dimension is already expected on the basis of M-theory. However, the
additional hidden timelike 12th dimension implied in $SO(c+1,1)$ is not
included in U-duality or in M-theory, it is a new feature discovered in the
context of S-theory.

\item  It is well known that the special form of $S_{\alpha \beta }=\gamma
_{\alpha \beta }^\mu \,\,p_\mu ,$ with the 11D momentum $p^2=0,$ corresponds
to a reference state (\ref{refstate}) that gives the 11D supergravity states 
$2_{bosons}^7+2_{fermions}^7.$ The fields corresponding to these states
depend on 11 momenta or coordinates (some of which may be compactified).
Their interactions is described by 11D supergravity, which follows by
requiring local supersymmetry. According to S-theory there are generalized
supergravity theories that follow from other special forms of $S.$ One such
example is \cite{ibstheory} 
\begin{equation}
S_{\alpha \beta }=\gamma _{\alpha \beta }^{MN}\,\left( p_Mp_N^{\prime
}-p_Np_M^{\prime }\right) ,\quad p^2=0,\,\quad p\cdot p^{\prime }=0,
\label{12dalg}
\end{equation}
where $p,p^{\prime }$ are 12D ``momenta". This special form of $S$ is related
to the 11D one by an $SO(10,2)$ boost that brings $p$ or $p^{\prime }$ to a
standard form. The supermultiplet based on this reference state is also the
11D supergravity multiplet, but this construction shows that this multiplet
is also a basis for 12D supertrasformations involving the new $SO(10,2)$
covariant superalgebra. Hence it implies that the 11D supergravity fields
can be extended to 12D by allowing them to depend on {\it a pair} of 12D
``momenta'' or coordinates, with the proper constraints 
\footnote{
After the talks that relate to the present paper and \cite{ibstheory},
the algebra (\ref{12dalg}) has been recognized to be at the basis of
new supersymmetric models as  
reported in \cite{martinec2} \cite{sezgin1}. In their case one of the
vectors $p,p^{\prime }$ is taken as constant, thus breaking the 12D
covariance. More generally, I emphasize that the superalgebra (\ref{12dalg})
permits these vectors to be dynamical operators. From this point of view the
models in \cite{martinec2} \cite{sezgin1} are constructed for a fixed
eigenvalue. Therefore I expect that there are generalizations of the models
in \cite{martinec2} \cite{sezgin1} that are fully 12D covariant. Generalized
models may be constructed by using a pair of ``momenta'' or coordinates.}.
All this implies that, among other things, there should exist a 12D
extension of supergravity of the type described here (possibly including
auxiliary non-propagating fields), as a prediction of S-theory. It is
therefore advisable to revive old attempts that almost succeeded to
construct an $SO(10,2)$ supergravity theory \cite{vannieu}. Apparently there
are renewed efforts in this direction.

\item  The examples above are special solutions of the condition $detS=0$
for the existence of BPS-like states. Other new solutions are provided in 
\cite{ibstheory}. The previous 12D example, as well as the other new
solutions, include central extensions that are Lorentz non-singlets. This
implies that the theory that can support such a superalgebra must include
p-branes. The classification of all solutions of $detS=0$ is tentamount to  a
classification of all the BPS-like states of the secret theory.
\end{itemize}

\section{Curved spacetime}

The examples above illustrate the methods of S-theory in flat spacetime. In
more complicated versions of the secret theory p-branes propagate in curved
compactified spaces (and/or non-compactified curved spacetime). Then from
the point of view of S-theory some or all of the 528 generators do not
commute with each other or with the fermionic generators (for example
momenta do not commute in curved spacetime). The structure of the
commutation rules is related to the geometry of the background. The
representation theory is technically harder and more interesting, but the
underlying physical concepts are similar to the flat case. Some non-Abelian
versions of S-theory would be related to more intricate contractions of
OSp(1/32). This can be achieved by different rescalings of various sets of
bosonic and fermionic generators, and then taking various contraction
limits. Details will be reported elsewhere \cite{ibentropy}.

The superalgebra may be further generalized by admitting fermionic central
extensions \cite{green}. In this case one may consider the contractions of
superagebras larger than OSp(1/32). A natural extension involves the 64
dimensional spinor space of 13D used in S-theory \cite{ibstheory} (thus
unifying types A,B superalgebras via a duality transformation involving the
13th dimension). One may then consider superalgebras with 64 fermionic
generators, such as OSp(1/64) or SU(1/32) contracted to the desired form, in
order to describe a further generalized version of S-theory in curved space,
now involving fermionic central extensions.

Recall that in the flat spacetime the Lorentz generators do not appear on
the right hand side of the anticommutator of the supercharges (i.e. they are
outside of the OSp(1/32) generators). When larger (contracted) superalgebras
are used for S-theory the Lorentz generators (or other isometries of the
curved spacetime) may be included as part of the enlarged algebra.

It is evident that there is much to be studied algebraically in S-theory
(and in its representations via field theoretic or string-like toy models) 
in order to learn the global
properties of the secret theory. In this talk I have described just the
beginning.

\end{document}